\documentclass[pra,twocolumn,10pt,aps,superscriptaddress,longbibliography]{revtex4-2}
\setcitestyle{square}
\usepackage{graphicx}  
\usepackage{float}
\usepackage{dcolumn}   
\usepackage{bm}        
\usepackage{amssymb}   
\usepackage{amsmath}
\usepackage{isomath}
\usepackage[usenames, dvipsnames]{color}
\usepackage{soul}
\usepackage{physics}
\usepackage[normalem]{ulem}
\usepackage{xpatch}
\usepackage{IEEEtrantools}
\usepackage[newcommands]{ragged2e}
\usepackage{subcaption}
\usepackage{pgffor}
\usepackage{etoolbox}
\usepackage{xprintlen}
\usepackage{comment}

\usepackage{xcolor}
\usepackage{hyperref}
\hypersetup{
     colorlinks  = true,
     linkcolor    = red,
     citecolor    = cyan,
     urlcolor     = magenta
     }

\begin{document}

\preprint{APS/123-QED}

\title{Nonlinear Corner States in Topologically Nontrivial Kagome Lattice}

\author{K Prabith}
\affiliation{Department of Aerospace Engineering, Indian Institute of Science, Bangalore 560012, India}

\author{Georgios Theocharis}
\affiliation{LAUM, CNRS-UMR 6613, Le Mans Universit\'{e}, Avenue Olivier Messiaen, 72085 Le Mans, France}

\author{Rajesh Chaunsali}%
 \email{Corresponding author: rchaunsali@iisc.ac.in}
\affiliation{Department of Aerospace Engineering, Indian Institute of Science, Bangalore 560012, India}%



\date{\today}

\begin{abstract}
We investigate a higher-order topological insulator (HOTI) under strong nonlinearity, focusing on the existence and stability of high-amplitude corner states, which can find applications in optics, acoustics, elastodynamics, and other wave-based systems. Our study centers on a breathing Kagome lattice composed of point masses and springs known to exhibit edge and corner states in its linear regime. By introducing onsite cubic nonlinearity, we analyze its impact on both edge and corner states. The nonlinear continuation of the corner state unveils stable high-amplitude corner states within the lattice, featuring non-zero displacements at even sites from the corner -- a characteristic absent in the linear limit. Interestingly, the nonlinear continuation of the edge state reveals its transformation into distinct families of high-amplitude corner states via two pitchfork bifurcations. While some states maintain stability, others become unstable through real instability and Neimark-Sacker bifurcation. These unstable corner states dissipate their energy into the edges and the bulk over an extended period, as corroborated by long-time dynamical simulations. Consequently, our study provides insights into achieving significant energy localization at the corners of HOTIs through various classes of nonlinear states.
\end{abstract}

\maketitle

\section{\label{Sec1}INTRODUCTION\protect\\}
In recent years, higher-order topological insulators (HOTIs) have emerged as a novel category of topological materials, distinguished by the emergence of topologically protected, gapless states localized at the corners or hinges of 2D and 3D lattices~\cite{benalcazar2017quantized, benalcazar2017electric, schindler2018higher, peterson2018quantized, serra2018observation}. Initially, this class included second-order topological quadrupole insulators featuring nonzero quadrupole moments, necessitating negative hopping. However, an alternative approach, rooted in nonzero bulk dipole moments and exploiting crystalline symmetry, has been proposed to realize second-order topological insulators~\cite{BenalcazarCn2019}. This method utilizes the ``filling anomaly" to establish the presence of higher-order states. The exploration of HOTIs has yielded a myriad of examples in both classical and quantum domains, sparking innovations in photonics~\cite{medina2023influence,medina2023corner}, plasmonics~\cite{proctor2021higher}, phononics~\cite{ma2023tuning,zhou2023visualization}, magnets~\cite{	yin2022topological}, stochastic systems~\cite{tang2021topology} and circuit design~\cite{ezawa2019non,yang2020observation} (see also the review by Xie et al.~\cite{xie2021higher} and references therein).

The entire framework of higher-order topology rests on the linear dynamics of the lattice. Exploring the interplay between nonlinearity and higher-order topology is one of the emerging research questions. Recently, this interplay has been investigated in the context of conventional (not higher-order) topological systems, revealing several intriguing phenomena~\cite{smirnova2020nonlinear}. For example, nonlinearity has been utilized to tune the existence and stability of topological edge and interface states~\cite{ablowitz2014linear, leykam2016edge, kartashov2016modulational, dobrykh2018nonlinear, pal2018amplitude, vila2019role, snee2019edge, darabi2019tunable, mukherjee2021observation, chaunsali2021stability, ma2021topological, tempelman2021topological, kartashov2022observation, jezequel2022nonlinear, li2023topological, michen2023nonlinear, rosa2023amplitude}, and to generate harmonics~\cite{kruk2019nonlinear, wang2019topologically, zhou2020switchable}. Furthermore, insights from topological band theory have revealed special characteristics of bulk solitons and breathers~\cite{lumer2013self, solnyshkov2017chirality, marzuola2019bulk, mukherjee2020observation}. Nonlinear bulk solutions have also been used to interpret nonlinear edge solutions~\cite{smirnova2019topological, chaunsali2023dirac}. Finally, through some special types of nonlinearities, ``self-induced" boundary states~\cite{hadad2016self, savelev2018topological, chaunsali2019self, d2019nonlinear, maczewsky2020nonlinearity, liu2022self, zhou2022topological} and domain walls~\cite{chen2014nonlinear, hadad2017solitons, poddubny2018ring, ma2023nonlinear} have been observed. However, there are very limited studies on nonlinear HOTIs~\cite{zangeneh2019nonlinear, tao2020hinge, zhang2020nonlinear, kirsch2021nonlinear, kompanets2024observation}. In particular, the existence and stability of different classes of high-order topological states in strongly nonlinear lattices remain unexplored.

In this study, we investigate a breathing Kagome lattice composed of point masses and springs, which exhibits onsite cubic nonlinearity. This model, a system of second-order ordinary differential equations, is a universal model applicable not only in mechanical systems~\cite{allein2023strain} but also in optical~\cite{vanel2017asymptotic,palmer2022asymptotically}, electrical~\cite{lee2018topolectrical}, and superconducting circuits~\cite{lazarides2018superconducting}. On-site nonlinearity in mechanical settings is common due to geometric nonlinearity~\cite{remoissenet2013waves}, while in optical and superconducting (Josephson-Junctions) settings, it arises due to the Kerr effect~\cite{boyd2008nonlinear} and magnetic flux~\cite{remoissenet2013waves}, respectively. While Kagome lattices have been previously explored for bulk nonlinear breathers~\cite{law2009localized, zhu2010defect, vicencio2013discrete, molina2012localized}, recent studies have revealed the presence of higher-order topological states at the corners of Kagome lattices, resulting from topological transitions~\cite{ezawa2018higher, xue2019acoustic,ni2019observation,kempkes2019robust, wu2020plane,wang2021elastic,herrera2022corner}. Our focus is on probing the existence and stability of highly nonlinear states localized at the corners of the lattice. We employ a nonlinear continuation technique on topological corner states to obtain a family of high-amplitude corner states. Additionally, since the Kagome lattice also supports edge states, we utilize a nonlinear continuation technique on these states to reveal a whole new family of high-amplitude corner states, as depicted by detailed bifurcation diagrams.

\section{\label{Sec2}SYSTEM AND ITS LINEAR DYNAMICS\protect\\}
\begin{figure}[!t]
    \includegraphics[width=1\linewidth]{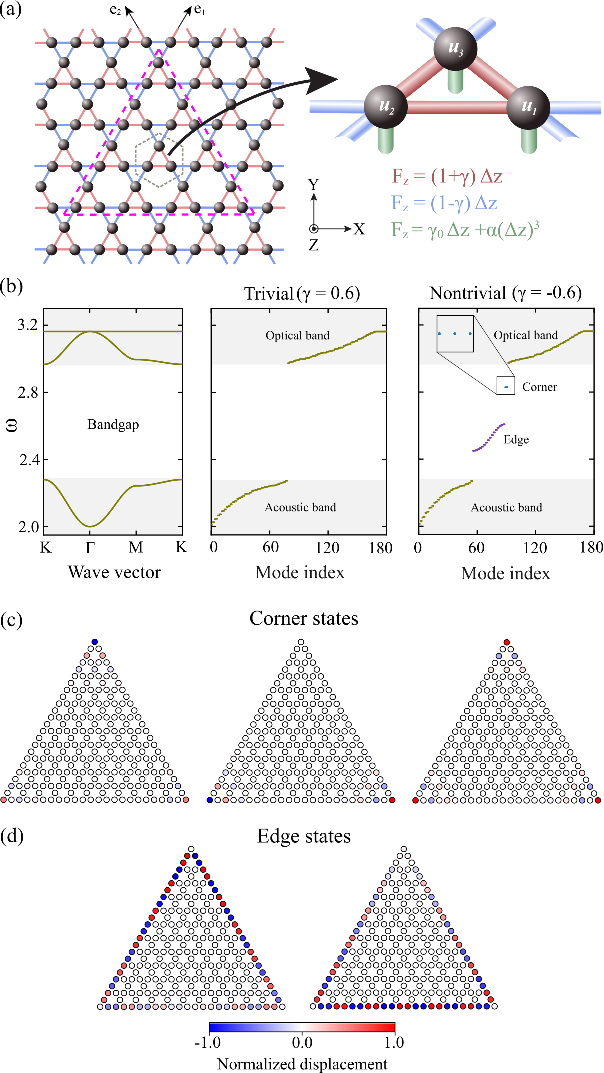}
    \caption{\label{Fig1}Nonlinear Kagome lattice and its linear characteristics: (a) Schematic of a Kagome lattice composed of point masses connected by linear springs (red and blue) and grounded with nonlinear spring (green). The inset displays the unit cell in 3D. The expressions for the force ($F_z$) with respect to the deformation ($\Delta z$) are indicated using different colors. (b) Dispersion diagram and eigenspectra of the finite triangular lattice (shown by the pink dashed line) under trivial ($\gamma = 0.6$) and nontrivial ($\gamma = -0.6$) conditions. Corner and edge states are highlighted inside the bandgap. (c) \& (d) Profiles of corner and edge states, respectively. The color bar denotes the normalized displacement of masses in the out-of-plane direction.}
\end{figure}

We consider an infinite Kagome lattice composed of point masses and springs, as illustrated in Fig.~\ref{Fig1}a. The unit cell (inset) comprises three masses connected by intracell stiffness, $k_1$ (highlighted in red), and intercell stiffness, $k_2$ (highlighted in blue). The masses are secured via a grounded spring (depicted in green), characterized by linear stiffness $\gamma_0$, and a parameter $\alpha$ introducing cubic nonlinearity. We consider one degree of freedom per particle, which could potentially represent either out-of-plane motion (transverse vibrations) in a lattice of masses~\cite{duan2023numerical} or rotations in a lattice of spinners~\cite{vila2019role,fang2023dispersion} in a mechanical context. Consequently, the nondimensionalized equations of motion for the masses within the unit cell, neglecting any dissipative effects, are expressed as follows:
\begin{eqnarray}
\ddot{u}_{1_{m,n}} + k_1(2{u}_{1_{m,n}} - {u}_{2_{m,n}} - {u}_{3_{m,n}}) + \gamma_0 {u}_{1_{m,n}} + \alpha {u}^3_{1_{m,n}}\nonumber \\
 + k_2(2{u}_{1_{m,n}}-{u}_{2_{m+1,n-1}} - {u}_{3_{m,n-1}})  = 0\nonumber\\
\ddot{u}_{2_{m,n}} + k_1(2{u}_{2_{m,n}}-{u}_{1_{m,n}} - {u}_{3_{m,n}}) + \gamma_0 {u}_{2_{m,n}} + \alpha {u}^3_{2_{m,n}}\nonumber \\
+ k_2(2{u}_{2_{m,n}}-{u}_{1_{m-1,n+1}} - {u}_{3_{m-1,n}})  = 0\nonumber\\
\ddot{u}_{3_{m,n}} + k_1(2{u}_{3_{m,n}}-{u}_{1_{m,n}} - {u}_{2_{m,n}}) + \gamma_0 {u}_{3_{m,n}} + \alpha {u}^3_{3_{m,n}}\nonumber \\
+ k_2(2{u}_{3_{m,n}}-{u}_{1_{m,n+1}} - {u}_{2_{m+1,n}})  = 0\nonumber\\  
\label{Eq1}
\end{eqnarray}
where the variables ${u}_{1_{m,n}}$, ${u}_{2_{m,n}}$, and ${u}_{3_{m,n}}$ represent the normalized out-of-plane displacements of the three masses in the unit cell. The indices $m$ and $n$ indicate the unit cell's position along its two basis vectors $e_1$ and $e_2$, as depicted in Fig.~\ref{Fig1}a. Overdots denote derivatives with respect to time. The stiffness quantities $k_1$ and $k_2$ are parameterized by a scalar $\gamma$ such that $k_1 = 1 + \gamma$ and $k_2 = 1 - \gamma$, where $\gamma \in (-1,1)$.

In the linear limit ($\alpha \rightarrow 0$), assuming a harmonic plane-wave solution, we obtain the dispersion diagram for the parameters $\gamma = \pm 0.6$ and $\gamma_0 = 4$, as depicted in Fig.~\ref{Fig1}b. It consists of two dispersive bands and one flat band, with the latter touching the upper dispersive band at the $\mathrm{\Gamma}$ point. For nonzero $\gamma$, a band gap exists between the two dispersive bands, and the system is a \textit{nontrivial} HOTI for $\gamma < -1/3$~\cite{ezawa2018higher, xue2019acoustic}. It can support higher-order topological states inside the band gap for finite lattices with triangle or parallelogram shapes.

In this analysis, we employ a finite Kagome lattice of triangular shape, cut from the infinite lattice (pink dashed triangle in Fig.~\ref{Fig1}a) with its boundaries fixed. The eigenspectra of the lattice, consisting of 78 unit cells, are presented for both the \textit{nontrivial} case ($\gamma = -0.6$) and the \textit{trivial} case ($\gamma = 0.6$) in Fig.~\ref{Fig1}b (refer to Appendix~\ref{appA} for the evolution of the spectrum as $\gamma$ is varied). In the \textit{trivial} lattice, no states exist within the band gap, while in the \textit{nontrivial} case, multiple states are observed within the band gap. These states are localized either at the edges or at the corners of the finite lattice, referred to as `edge states' and `corner states,' respectively, throughout the remainder of the manuscript.

The corner states are nearly degenerate~\cite{herrera2022corner} at the frequency $\sqrt{4+\gamma_0}$  and located at three corners of the finite triangular lattice, as shown in Fig.~\ref{Fig1}c. One unique characteristic of these corner states is that every alternate (even) site from the corner exhibits zero displacements~\cite{xue2019acoustic}. Meanwhile, the profiles of the two edge states corresponding to the two highest eigenfrequencies in the edge spectrum are displayed in Fig.~\ref{Fig1}d. Interestingly, these two edge states have nearly the same eigenfrequencies. The first edge state exhibits maximum displacements localized at the left and right edges of the triangle, while the latter has them at the bottom edge. In the following sections, we will not only focus on how nonlinearity affects the corner states but also explore its impact on edge states, which may lead to the emergence of new classes of corner states in nonlinear lattices.

\section{\label{Sec3}NONLINEAR CONTINUATION\protect\\} 
In this analysis, we apply a strong hardening nonlinearity with $\alpha = 0.8$ to the grounded springs of the Kagome lattice to investigate its influence on corner and edge states. We utilize a Newton solver to numerically obtain nonlinear periodic solutions at various frequencies and lattice energies, thereby revealing the nonlinear normal modes (NNMs). For clarity, we term NNMs obtained by the nonlinear continuation of corner and edge states as `$\mathrm{NNM_c}$' and `$\mathrm{NNM_e}$', respectively, in subsequent sections. Additionally, we determine the linear stability of these NNMs using Floquet theory and analyze the nature of instabilities through the examination of Floquet multipliers (FMs).

\subsection{Continuation of corner state}
Since we had observed three nearly degenerate corner states in Fig.~\ref{Fig1}c, we now prepare a new corner state localized at only one (top) corner of the lattice through linear superposition. This state is then provided as the initial condition to the Newton solver. We make this choice for simplicity as we investigate a family of high-amplitude corner states at only one corner region. It is important to note that this approach remains valid for a large lattice, where coupling effects between states localized at different corners are negligible. 

\begin{figure}[!t]
\includegraphics[width=1\linewidth]{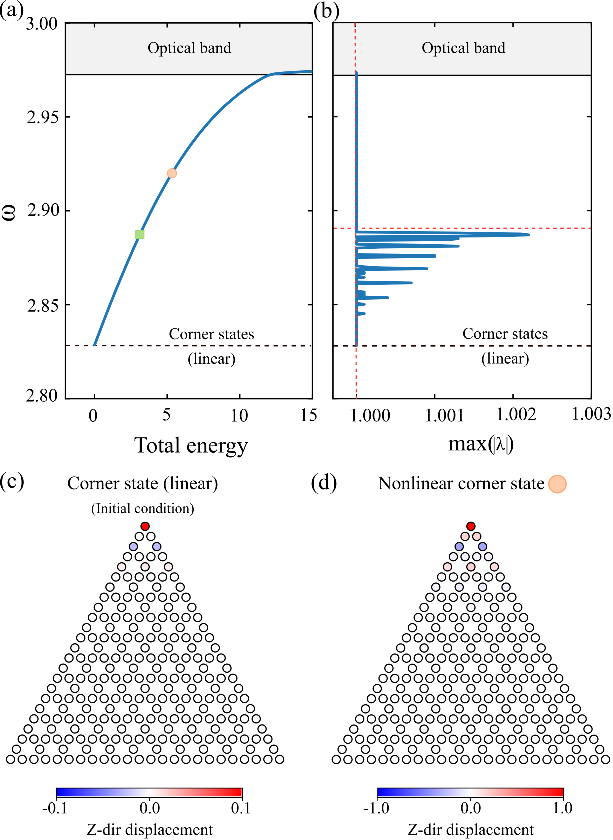}
\caption{\label{Fig2} Nonlinear continuation of the corner state: (a) Frequency-energy plot showing the evolution of $\mathrm{NNM_c}$ with lattice energy. (b) Maximum absolute value of Floquet multipliers `$\lambda$' describing the stability of periodic solutions, becoming unstable when they exceed unity (vertical dashed line). The red horizontal dashed line indicates the beginning of the stable region. (c) The initial condition for numerical continuation obtained by the superposition of three corner states in the linear limit. (d) Profile of the $\mathrm{NNM_c}$ measured at $\omega = 2.92$, as indicated by the circular marker in (a). It shows a high-amplitude corner state. The color bar represents the out-of-plane displacements of the masses.}
\end{figure}

Figure~\ref{Fig2}a displays a frequency-energy plot illustrating the evolution of $\mathrm{NNM_c}$ as the lattice energy in the system increases. The lattice energy represents the total energy stored in the lattice due to the deformation of springs and the displacement of the masses from their equilibrium positions. We observe that the frequency of the $\mathrm{NNM_c}$ increases with the total energy in the system due to the hardening nature of nonlinearity, as observed in a one-dimensional nonlinear chain~\cite{chaunsali2021stability, tempelman2021topological}. Moreover, they enter the optical band at higher energy levels and resonate with bulk states, resulting in a decrease in slope as the total energy rises.

\begin{figure}[!t]
\includegraphics[width=1\linewidth]{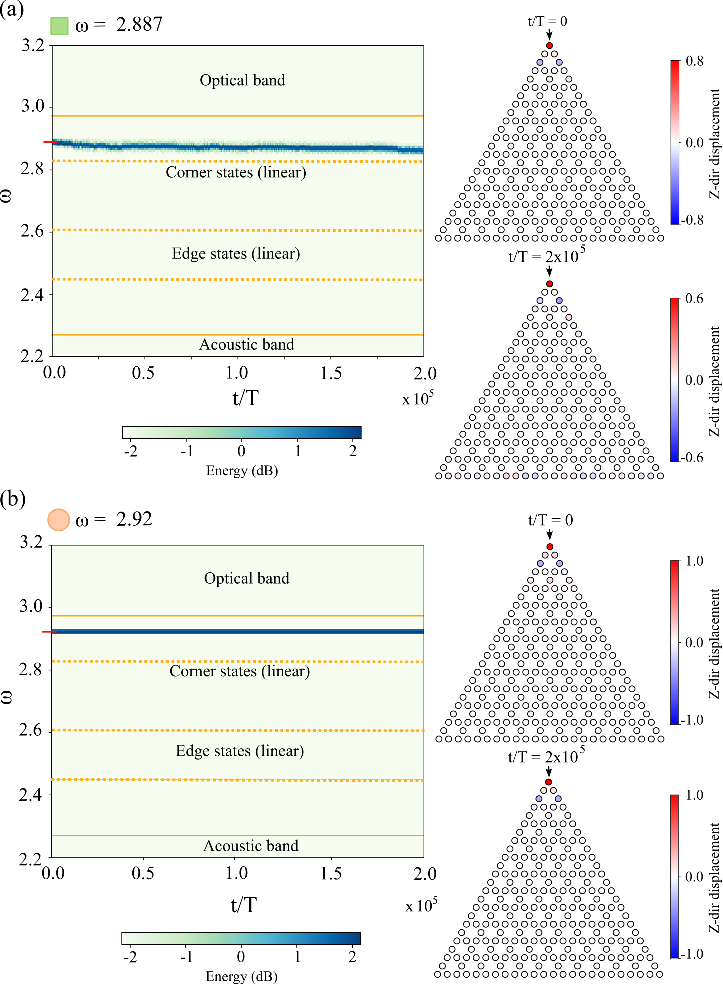}
\caption{\label{Fig3} Transient analysis of $\mathrm{NNM_c}$ to demonstrate its stability nature: (a) Short-time Fourier transform (STFT) of the displacement measured at the top particle (marked with a black arrow) for a frequency of $\omega = 2.887$. The instability in this state results in a gradual downshift of frequency towards a new state over time. (b) STFT of the displacement measured at the same particle for a frequency of $\omega = 2.92$, maintaining the same state over an extended period, indicating its stable nature. The color bar represents the total energy in the system, expressed in decibels (dB). Profiles at the initial time ($t = 0$) and the final time ($t = 2 \times 10^5T$) are also included for both (a) and (b).}
\end{figure}

The linear stability of $\mathrm{NNM_c}$ varies within the bandgap, as depicted in Fig.~\ref{Fig2}b, where the maximum amplitude of FMs is plotted as a function of frequency. It is well-known that periodic solutions remain linearly stable when the maximum amplitude of FMs is less than or equal to unity, but instability arises when it exceeds unity. Initially, the $\mathrm{NNM_c}$ remain stable up to $\omega = 2.84$ due to weak nonlinear effects. In this range, the profile of the $\mathrm{NNM_c}$ is similar to that of a corner state in the linear limit, as shown in Fig.~\ref{Fig2}c. However, beyond $\omega = 2.84$, the $\mathrm{NNM_c}$ become unstable until about $\omega = 2.89$, as depicted in Fig.~\ref{Fig2}b. It is worth noting that the magnitude of instability is very small, and it even approaches unity at some intermediate frequencies within this range. These are finite-size instabilities~\cite{marin1998finite}, which are expected to reduce further as the size of the system increases (see Appendix~\ref{appB} for more details). The profile in this range appears to be localized around the corner of the lattice; hence, the $\mathrm{NNM_c}$ can also be referred to as a nonlinear corner state. Interestingly, for $\omega > 2.89$, $\mathrm{NNM_c}$ becomes stable inside the bandgap with a high amplitude of vibration. Figure~\ref{Fig2}d shows one such stable high-amplitude corner state at $\omega = 2.92$. Importantly, these high-amplitude corner states exhibit non-zero displacements at even sites from the top corner, a feature that distinguishes them from corner states in the linear limit.

To demonstrate the linear stability of  $\mathrm{NNM_c}$, we conduct long-time transient simulations for a duration of $2 \times 10^5 T$, where $T$ represents the time period of $\mathrm{NNM_c}$. At first, the  $\mathrm{NNM_c}$ at $\omega = 2.887$ is used as the initial condition, with the addition of white noise at an amplitude of 1 \% of the displacement. Figure~\ref{Fig3}a presents the short-time Fourier transform (STFT) obtained from the time history of the top particle, marked with a black arrow. Since this $\mathrm{NNM_c}$ is linearly unstable, it deviates from the initial frequency and slowly moves towards a low-frequency state that is stable. The spatial profile at $t/T = 2\times10^5$ is also depicted in Fig.~\ref{Fig3}a, resembling the $\mathrm{NNM_c}$ at $\omega = 2.863$, which falls inside the small stable region within the frequency range of 2.84-2.89.

Now we present the STFT of the same particle when the $\mathrm{NNM_c}$ at $\omega$ = 2.92 is given as the initial condition. Since this state is linearly stable, from Fig.~\ref{Fig3}b, it is evident that the frequency of the state remains constant and maintains the initial state over an extended period. Therefore, we have verified the existence of stable high-amplitude corner states within the bandgap.

\subsection{Continuation of edge state}
\begin{figure*}[!t]
\includegraphics[width=1\linewidth]{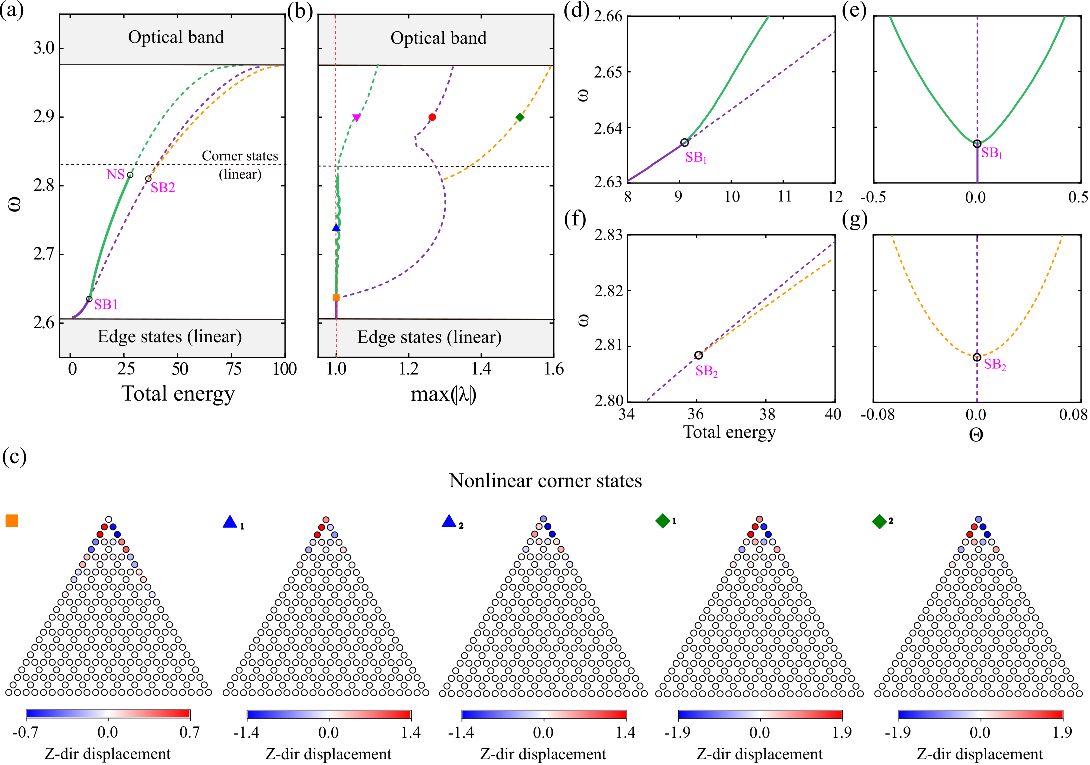}
\caption{\label{Fig4} Nonlinear continuation of the edge state: (a) Frequency-energy plot showing the evolution of $\mathrm{NNM_e}$ with lattice energy. It splits into different branches as the energy in the system increases, depicted using different colors. The bifurcation points on these branches are labeled as $\mathrm{SB_1}$, $\mathrm{SB_2}$, and $\mathrm{NS}$. (b) Maximum absolute value of Floquet multipliers `$\lambda$' as a function of $\omega$, indicating the stability of periodic solutions. The stable solutions are shown by thick lines, while unstable solutions are indicated by dashed lines. (c) The profiles of various $\mathrm{NNM_e}$, measured at the frequencies indicated by different markers in (b), show distinct states localized at the corner of the lattice. The color bar represents the out-of-plane displacements of the masses. (d)-(g) Zoomed sections of the energy plot and asymmetry coefficient `$\mathrm{\Theta}$' varied with respect to frequency around $\mathrm{SB_1}$ and $\mathrm{SB_2}$ respectively. Here, the color bar denotes the normalized displacement of masses in the out-of-plane direction.}
\end{figure*}

We use an edge state, depicted in the first part of Fig.~\ref{Fig1}d, as the initial condition for the Newton solver. The frequency-energy plot and the stability diagram of the $\mathrm{NNM_e}$ are illustrated in Figs.~\ref{Fig4}a-b. Interestingly, as the total energy increases, the edge state transforms into various types of states localized at the corner of the lattice. This transformation is evident from their profiles, as depicted in Fig.~\ref{Fig4}c. Hence, for this particular edge state, the $\mathrm{NNM_e}$ can also be referred to as a nonlinear corner state. At lower, intermediate, and higher frequencies, we observe one, three, and five types of nonlinear corner states, respectively. The differences in these nonlinear corner states can be explained using the asymmetry coefficient, $\mathrm{\Theta}$, defined as~\cite{li2021existence}.
\begin{equation}
    \mathrm{\Theta} = \frac{E_1-E_2}{E_1+E_2}, 
    \label{Eq3}
\end{equation}
where $E_1$ and $E_2$ represent the potential energies stored in the grounded springs undergoing deformations in positive and negative Z-directions, respectively. It satisfies the relation $E = E_1 + E_2$, where $E$ is the total potential energy stored in the grounded springs. Therefore, for a particular nonlinear corner state, $\mathrm{\Theta}$ examines the symmetry in energy distribution about the X-Y plane.

As depicted in Figs.~\ref{Fig4}a-b, the frequency of the $\mathrm{NNM_e}$ increases, and they undergo different bifurcations as the lattice energy is increased. The $\mathrm{NNM_e}$ remains stable until $\omega = 2.637$ (marked by the thick purple line), although minor finite-size instabilities persist within this range. The profile just below a frequency of $\omega = 2.637$ is shown in Fig.~\ref{Fig4}c, indicated by a square marker. It reveals an entirely new class of corner states, distinct from the $\mathrm{NNM_c}$ discussed in the previous section. A zero displacement occurs at the corner mass, and the displacements on the left edge are in anti-phase with those on the right edge. This illustrates the symmetric nature ($\mathrm{\Theta}=0$) of the state about the X-Y plane.

Increasing further the energy, a bifurcation occurs at $\omega = 2.637$, leading to the separation of the stable symmetric branch into an unstable symmetric branch (indicated by the dashed purple line) and two stable asymmetric branches (represented by the thick green line). As the energies of both asymmetric branches are equal, they overlap in the energy plot. This bifurcation point is a symmetry-breaking point, denoted as `$\mathrm{SB_1}$' in subsequent diagrams. Figures~\ref{Fig4}d-e illustrate zoomed sections of the energy plot and the asymmetry coefficient plotted around the point $\mathrm{SB_1}$. In Fig.~\ref{Fig4}e, two distinct curves are presented for the stable asymmetric branches since the signs of $\mathrm{\Theta}$ are opposite to each other. They clearly show a pitchfork bifurcation occurring at $\mathrm{SB_1}$, resulting in two stable states and one unstable state beyond $\mathrm{SB_1}$.

\begin{figure*}[!t]
\includegraphics[width=1\linewidth]{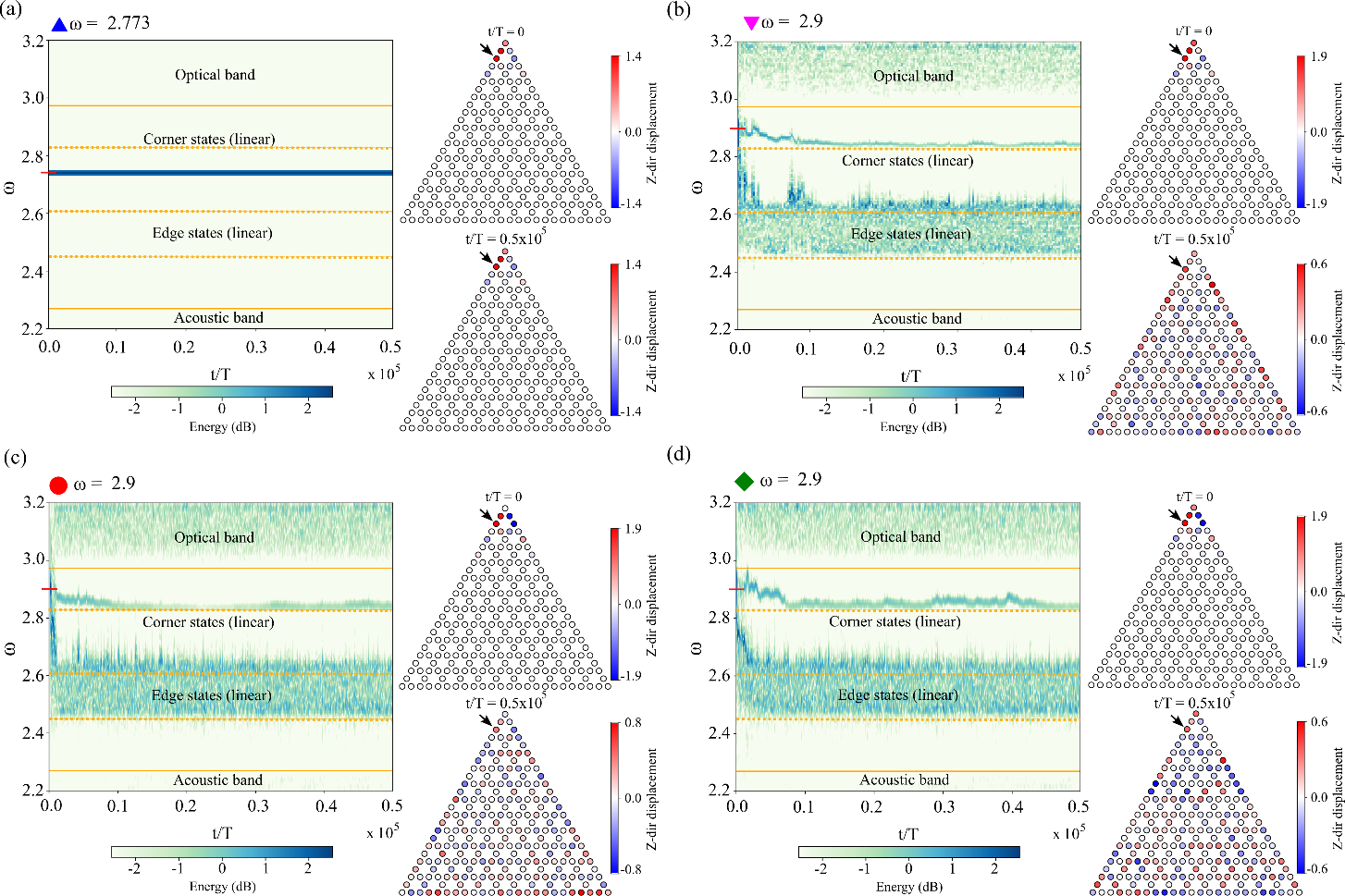}
\caption{\label{Fig5} Transient analysis of $\mathrm{NNM_e}$ to demonstrate its stability nature: (a) Short-time Fourier transform of the displacement measured at a specific particle (marked with a black arrow) on the stable $\mathrm{NNM_e}$, indicated using a triangle marker in Fig.~\ref{Fig4}. The frequency remains constant, and the $\mathrm{NNM_e}$ maintains its shape throughout the period. (b)-(d) STFT of the displacement measured at the same particle, but for unstable $\mathrm{NNM_e}$ highlighted in Fig.~\ref{Fig4}.  It deviates from the initial state quickly since the magnitude of instability is high. The final state contains the frequencies of linear corner states, edge states, and bulk states. The color bar represents the total energy in the system, expressed in decibels (dB). Profiles at the initial time ($t = 0$) and the final time ($t = 0.5\times10^5T$) are also included. Here, the color bar denotes the normalized displacement of masses in the out-of-plane direction.}
\end{figure*}
Upon tracing these stable branches, we observe that the magnitude of the asymmetry coefficient increases with frequency and exhibits strong asymmetry at large $\omega$. This is evident from their profiles, highlighted with triangular markers in Fig.~\ref{Fig4}c. We now observe a nonzero displacement at the top corner mass, and the displacements at the left and right edges differ in magnitude and phase. Hence, this represents a new type of stable nonlinear corner states. With the increase in energy, finite instabilities become more significant until a frequency of $\omega = 2.816$. However, the magnitude of such instabilities is small compared to other types of instabilities. Therefore, stable high-amplitude corner states can exist in this frequency range when a large triangular lattice is constructed.

Beyond $\omega = 2.816$, a qualitative change in the $\mathrm{NNM_e}$ occurs, wherein the magnitude of instability starts to increase rapidly (represented by a dashed green line). While analyzing the FMs, we observe that a pair of Floquet multipliers begins to leave the unit circle along the complex plane as the energy is increased (see Appendix~\ref{appB} for more details). This is an oscillatory instability and occurs due to Neimark-Sacker bifurcation at $\omega = 2.816$, denoted by `$\mathrm{NS}$' in the plot. Therefore, the $\mathrm{NNM_e}$ appearing beyond $\mathrm{NS}$ are largely unstable. Furthermore, this unstable branch enters the optical band at high-energy levels and resonates with the bulk states.

Following the unstable symmetric branch beyond $\mathrm{SB_1}$, we find that the FMs leave the unit circle along the real axis. This represents a real instability, which increases rapidly with increasing energy. Throughout this branch, the nonlinear corner states maintain their symmetry. With further increase in energy, a second symmetry-breaking event ($\mathrm{SB_2}$) occurs at $\omega = 2.808$. The unstable symmetric branch then splits into two unstable asymmetric branches (represented by the dashed orange line) and an unstable symmetric branch (indicated by the dashed purple line). Similar to the previous bifurcation, the asymmetric branches overlap in the energy plot and present two distinct curves in the asymmetry plot, as shown in Figs.~\ref{Fig4}f-g. It exhibits a pitchfork-like structure but with all unstable branches. The profiles on these unstable asymmetric branches are plotted in Fig.~\ref{Fig4}c, with diamond markers. Compared to the symmetric nonlinear corner state, these new states exhibit a nonzero displacement at the top corner mass, while the remaining masses have the same displacements. Hence, the value of the asymmetry coefficient is very small, as seen in Fig.~\ref{Fig4}g. Moreover, these asymmetric corner states are in anti-phase with each other at the top corner mass, vibrating in opposite directions at the same frequency. 

To explore the evolution of these $\mathrm{NNM_e}$ over time, a long-time transient simulation is carried out for $t = 0.5 \times 10^5T$. The STFT obtained from the time history of a specific particle (marked with a black arrow) is provided in Figs.~\ref{Fig5}a-d. For the stable $\mathrm{NNM_e}$, highlighted with a triangle marker in Fig.~\ref{Fig4}c, the frequency of the state remains constant and maintains its shape over time. However, for unstable $\mathrm{NNM_e}$, the frequency quickly deviates from that of the initial state due to the high magnitude of instability. Moreover, the final state contains frequencies corresponding to linear corner states, edge states, and bulk states, resulting in a complex profile. Thus, among the five types of corner states obtained through the continuation of the edge state, only two types of nonlinear corner states (marked with triangles) remain stable at high frequencies.

\section{CONCLUSIONS\protect\\}
In this paper, we investigate a HOTI composed of a Kagome lattice that supports edge and corner states in the linear limit. We examine the existence and stability of high-amplitude corner states resulting from cubic on-site nonlinearity. We perform nonlinear continuation for both the corner state and the edge state to obtain their NNMs. Our findings indicate that the NNMs from the continuation of corner state can be linearly stable in a specific frequency regime, hinting at robust and high-amplitude energy localization at the corners. Additionally, we observe different families of high-amplitude corner states when an edge state is continued into the nonlinear regime. This phenomenon occurs through two pitchfork bifurcations, transforming a symmetric corner state into two asymmetric corner states and one symmetric corner state. Some of them remain stable, while others lose their stability through real instability and Neimark-Sacker bifurcation.

The key finding of this work is the possibility of having different classes of high-amplitude corner states bifurcated from the edge state under nonlinear conditions. This phenomenon depends on the type of nonlinearity in the lattice. Therefore, investigating the role of inter-site nonlinearity and different forms of nonlinearity on the bifurcation and stability nature of high-amplitude corner states will be an interesting avenue for future research. Moreover, exploring the effect of non-Hermiticity~\cite{manda2024skin} on the nonlinear characteristics of such topological states would be interesting.

\begin{acknowledgments}
R.C. acknowledges the funding support by the Science and Engineering Research Board (SERB), India, through the Start-up Research Grant SRG/2022/00166.
\end{acknowledgments}

\appendix 
\section{Eigenspectrum of the finite Kagome lattice as a function of $\gamma$} \label{appA}
\begin{figure}[!h]
\includegraphics[width=1\linewidth]{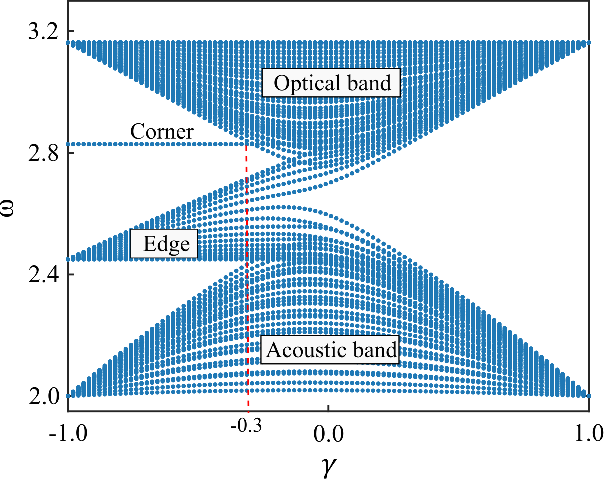}
\caption{\label{Fig6} Eigenspectrum for a triangular-shaped Kagomelattice plotted as a function of $\gamma$.}
\end{figure}

The eigenspectrum of the finite Kagome lattice of triangular shape for different values of $\gamma$, spanning between $-1$ and $1$, is shown in Fig.~\ref{Fig6}. Since the system is a \textit{nontrivial} HOTI for $\gamma < -0.3$, we observe the emergence of corner states inside the bandgap.

\section{Type of instabilities in the system} \label{appB}
Typically, the instabilities in the system are determined by analyzing the FMs of the NNMs. When we plot the FMs in the complex plane, they lie on a unit circle for a stable conservative system. Instability arises when some FMs start to leave the unit circle. This can occur in two ways: when two complex conjugate FMs collide on the real axis and leave the unit circle along the real axis, or when two pairs of complex conjugate FMs collide elsewhere on the unit circle and leave the unit circle as a quadruplet of FMs. The first type of instability is called \textit{real instability}, and the latter one is called \textit{oscillatory instability}.

\begin{figure*}[!t]
\includegraphics[width=1\linewidth]{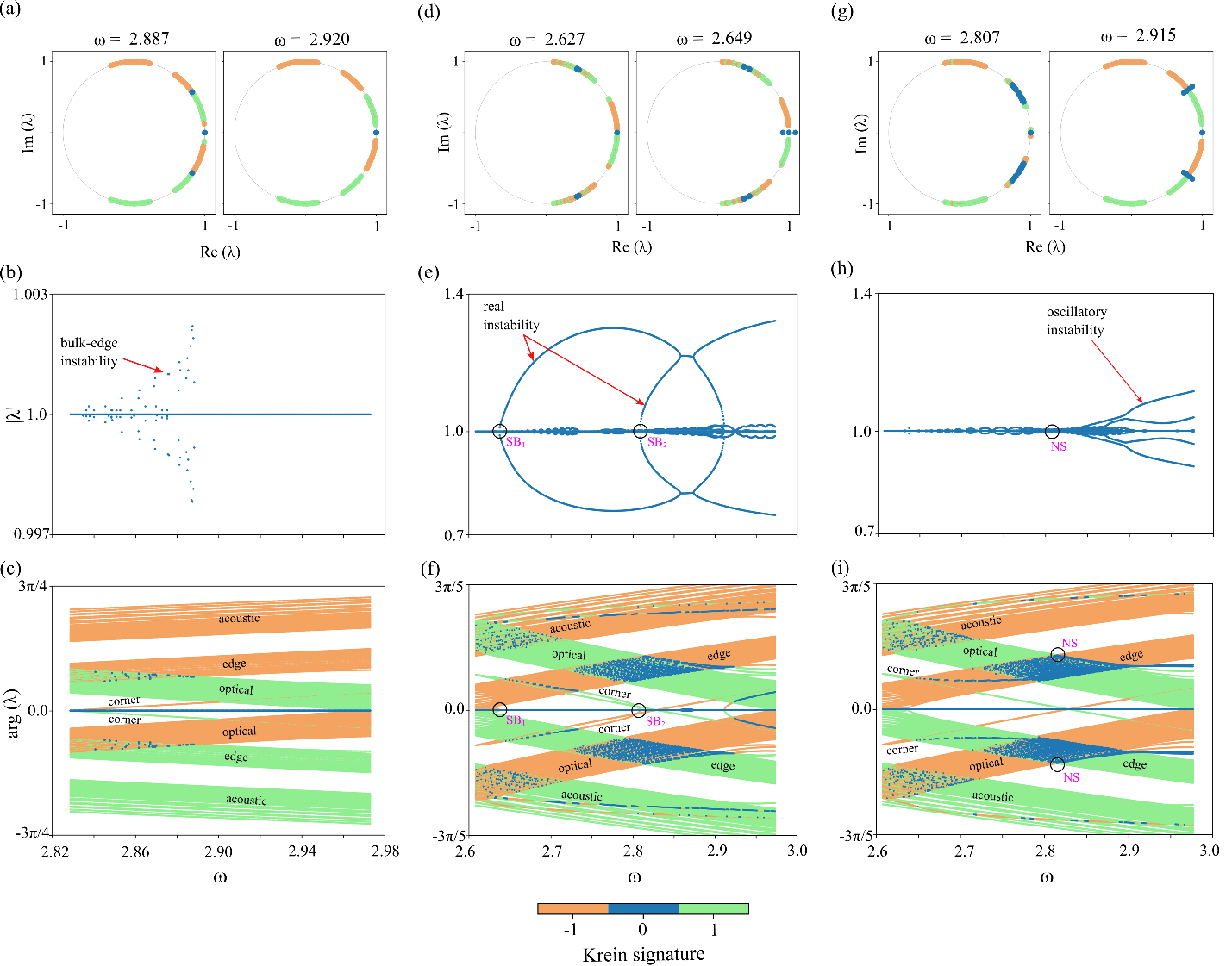}
\caption{\label{Fig7} Variation in FMs with an increase in frequency: (a) FMs at two specified frequencies are presented in the complex plane for NNMs obtained through the continuation of corner state. The color indicates the Krein signature of each FM. (b) Variations in the absolute value of FMs as the frequency increases. When $|\lambda|$ exceeds unity, the system becomes unstable. Types of instabilities are also indicated. (c) Variations in the phase of FMs with frequency. Blue dots represent FMs with zero Krein signature, indicating instability. [(d)-(i)] Similar plots are provided for the two branches of NNMs obtained through the continuation of edge state. The circle marker shows the bifurcation points. }
\end{figure*}
Now, we examine the types of instabilities that arise in our system. Figure~\ref{Fig7}a shows the FMs of $\mathrm{NNM_c}$, measured at $\omega = 2.887$ and $2.92$. Each FM is colored according to its Krein signature~\cite{aubry2006discrete,flach2008discrete,chung2020linear}, which takes either $+1$ or $-1$ for FMs lying on the unit circle, excluding the one at $(+1,0)$, where it is zero. FMs leaving the circle also have a zero Krein signature. In Fig.~\ref{Fig7}a, we observe that at $\omega = 2.887$, FMs with opposite Krein signatures (orange and green arcs) collide with each other and leave the unit circle with a zero Krein signature (blue). Since the collision occurs at a point other than the real axis, it represents an \textit{oscillatory instability}. Due to the very small magnitude of this instability, these FMs appear to lie on the unit circle. However, at $\omega = 2.92$, the FMs do not collide with each other, and there are no FMs with zero Krein signature outside the unit circle, indicating that the state is linearly stable.

To verify this further, we plot the variations in FMs as a function of frequency. Figures~\ref{Fig7}b and \ref{Fig7}c depict how the absolute value ($|\lambda|$) and argument ($\arg(\lambda)$) of FMs change with increasing frequency. Within the frequency range of $2.84-2.89$, $|\lambda|$ exceeds unity, and there exists a collision between spectral bands of opposite Krein signatures (orange and green). This collision induces instabilities in the system, characterized by a zero Krein signature (blue). We can identify states associated with each spectral band since they are related to the eigenspectrum of the linear system, as reported in Ref.~\cite{chaunsali2021stability}. In Fig.~\ref{Fig7}c, these states are marked on each spectral band. Since the collision occurs between the bulk and edge states, we term this instability the `bulk-edge instability.' Such instabilities decrease in strength as the size of the system increases and are expected to vanish in the infinite lattice limit~\cite{marin1998finite}. Hence, they are also called \textit{finite-size instabilities}. Beyond $\omega = 2.89$, no collisions of states (no blue dots) are observed, although the corner states intersect bulk states in this frequency range. Therefore, these states are stable, as presented in the main text.

We conduct a similar investigation for the $\mathrm{NNM_e}$. In Fig.~\ref{Fig7}d, the FMs of symmetric nonlinear corner states (represented by the purple line in Fig.~\ref{Fig4}a) are plotted in the complex plane for $\omega = 2.627$ and 2.649. Notably, at $\omega = 2.627$, some FMs with a zero Krein signature lie close to the unit circle, indicating \textit{finite-size instability}. In contrast, at $\omega$ = 2.649, an FM with a zero Krein signature lies outside the unit circle on the real axis, suggesting \textit{real instability}. Furthermore, we plot $|\lambda|$ and $arg(\lambda)$ as functions of frequency in Figs.~\ref{Fig7}e-f. Within the frequency range of $2.607-2.637$, collisions between two bulk states from acoustic and optical bands occur, resulting in finite-size instabilities of very small magnitude. At $\omega$ = 2.637, we observe two complex conjugate FMs colliding on the real axis. This collision point is $\mathrm{SB_1}$. It is a bifurcation point beyond which the FMs depart from the unit circle along the real axis. Subsequently, the magnitude of instability increases rapidly, confirming real instability. Moreover, a second bifurcation occurs at $\omega = 2.808$, where two complex conjugate FMs collide again on the real axis. This point is represented by $\mathrm{SB_2}$. In short, real instabilities dominate in the system beyond $\omega = 2.637$, although some finite-size instabilities also persist.

Figures~\ref{Fig7}g-i depict the FMs of asymmetric corner states (indicated by the green line in Fig.~\ref{Fig4}a) plotted against frequency. Finite-size instabilities are observed within the system up to a frequency of $\omega = 2.816$, after which oscillatory instabilities emerge. This is evident in Fig.~\ref{Fig7}g, where FMs with zero Krein signature are notably positioned outside the unit circle, particularly at $\omega = 2.915$. In Figs.~\ref{Fig7}h and \ref{Fig7}i, it becomes apparent that the onset of oscillatory instability occurs when an isolated FM emerges from a spectral band (marked with a circle). This marked point, representing the onset of oscillatory instability, is identified as a bifurcation point known as the Neimark-Sacker bifurcation ($\mathrm{NS}$). The magnitude of such instability increases with the increase in frequency.

\def\bibsection{\section*{References}}

\bibliography{biblio.bib}

\end{document}